\documentclass[12pt]{article}
\usepackage{fullpage,sak,latexsym,amsmath,amssymb,epsfig}

\newfontobj{\class}{\bf}

\renewcommand{\P}{{\bf P}}
\newcommand{\Poly}{{\rm Poly}}

\newcommand{\NumP}{{\#\P}}
\newcommand{\CeqP}{{{\bf C}_{=}\P}}
\newcommand{\CneqP}{{{\bf C}_{\neq}\P}}
\class{NP}
\class{PH}
\class{NQP}
\class{FP}
\class{BPP}
\class{BQP}
\class{EQP}
\class{AWPP}
\class{AAWPP}
\class{GUM}
\class{GAM}
\class{GUAM}
\class{PUM}
\class{PAM}
\class{PUAM}
\class{UP}
\class{AM}
\class{MA}
\class{PP}
\class{NC}
\class{AC}
\class{ACC}
\class{QNC}
\class{EQNC}
\class{NQNC}
\class{BQNC}
\class{QACC}
\class{QAC}
\class{NQAC}
\class{NQACC}
\class{TC}
\class{QTC}

\newcommand{\co}[1]{{\rm co}{#1}}

\newcommand{\rats}{{\mathbb Q}}

\newcommand{\map}[3]{{#1}:{#2}\rightarrow{#3}}

\newcommand{\ket}[1]{|{#1}\rangle}

\newcommand{\cC}{{\cal C}}

\newcommand{\adj}[1]{{{#1}^{\dagger}}}
\newcommand{\Mod}{{\rm Mod}}
\newcommand{\two}{\{0,1\}}

\newtheorem{definition}{Definition}[section]

\newtheorem{lemma}[definition]{Lemma}
\newtheorem{theorem}[definition]{Theorem}
\newtheorem{corollary}[definition]{Corollary}

\newenvironment{proof}{\noindent{\bf Proof.}}{\hfill$\Box$ \bigskip}
\newenvironment{proofof}[1]{\noindent{\bf Proof of
{#1}.}}{\hfill$\Box$ \bigskip}

\title{Bounds on the Power of Constant-Depth Quantum Circuits}

\author{S.~Fenner\footnote{Dept.~of CS and Eng.,
University of South Carolina, Columbia, SC 29208,
$\{$fenner$|$zhang29$\}$@cse.sc.edu}
\and 
F.~Green\footnote{Dept.~of Math and CS, 
Clark University, Worcester, MA 01610, fgreen@black.clarku.edu}
 \and 
S.~Homer\footnote{Computer Science Department, Boston University,
 Boston, MA 02215, homer@bu.edu}
\and 
Y.~Zhang\footnotemark[1]
}

\begin{document}

\maketitle

\begin{abstract}
We show that if a language is recognized within certain error bounds
by constant-depth quantum circuits over a finite family
of gates, then it is computable in (classical) polynomial time.  In
particular, our results imply
\[ \EQNC^0 \subseteq \P, \]
where $\EQNC^0$ is the constant-depth analog of the class $\EQP$.

On the other hand, we adapt and extend ideas of Terhal \& DiVincenzo
\cite{TD:constant-depth} to show that, for any family $\cal F$ of
quantum gates including Hadamard and CNOT gates, computing the acceptance
probabilities of depth-five circuits over $\cal F$ is just as hard as
computing these probabilities for arbitrary quantum circuits over
$\cal F$.  In particular, this implies that
\[ \NQNC^0 = \NQACC = \NQP = \co{\CeqP}, \]
where $\NQNC^0$ is the constant-depth analog of the class $\NQP$.
This essentially refutes a conjecture of Green et al.\ that $\NQACC
\subseteq \TC^0$ \cite{GHMP:QAC}.
\end{abstract}

\section{Introduction}
\label{sec:intro}

Quantum decoherence is a major obstacle to maintaining long quantum
computations.  The first working quantum computers will almost
certainly be limited to realizing shallow---i.e.,
small-depth---quantum circuits.  This dilemma has inspired much
theoretical interest in the capabilities of these circuits,
particularly circuits that have constant depth and polynomial size.

Recently, people have found that much can be done with $O(\log
n)$-depth circuits.  For example, Cleve \& Watrous were able to
approximate the Quantum Fourier Transform over modulus $2^n$ with
$O(\log n)$-depth circuits \cite{CW:fourier}.  Log-depth seems to
present a barrier for many computational problems, however; getting
significantly shallower circuits appears difficult if not
impossible---unless gates of unbounded width (i.e., number of qubits,
or fan-in) are allowed.  This has led to the study of constant-depth
quantum circuits that can contain certain classes of unbounded fan-in
gates.

There are a number of unbounded-width gate classes studied in the
literature, most being defined in analogy to classical Boolean gates.
The generalized Toffoli gate (see Section~\ref{sec:prelims-gates}) is
the quantum equivalent of the unbounded Boolean AND-gate.  Likewise,
there are quantum equivalents of Mod-gates and threshold gates.  One
particular quantum gate corresponds to something taken almost
completely for granted in Boolean circuits---fan-out.  A fan-out gate
copies the (classical) value of a qubit to several other qubits at
once.\footnote{There is no violation of the No-Cloning Theorem here;
only the classical value is copied.}  Using these gates, one can
define quantum versions of various classical circuit classes: $\QNC^k$
(Moore \& Nilsson \cite{MN:quantum}), $\QAC^k$ and $\QACC^k$ (Moore
\cite{Moore:fanout}, Green et al.\ \cite{GHMP:QAC}), and $\QTC^k$
are analogous to $\NC^k$, $\AC^k$, $\ACC$, and $\TC^k$, respectively.
The case of particular interest is when $k=0$.  All these classes are
allowed constant-width gates drawn from a finite family.  The classes
differ in the additional gates allowed.  $\QNC^0$ is the most
restrictive class; all gates must have bounded width.  $\QAC^k$
circuits are allowed generalized Toffoli gates, and $\QACC^k$ circuits
are allowed $\Mod_q$-gates, where $q$ is kept constant in each circuit
family.  $\QTC^k$ circuits are allowed quantum threshold gates.  See
Section~\ref{sec:prelims-gates} for detailed definitions of most of
these classes.

Although quantum classes are defined analogously to Boolean classes,
their properties have turned out to be quite different from their
classical versions.  A simple observation of Moore \cite{Moore:fanout}
shows that the $n$-qubit fan-out gate and the $n$-qubit parity
($\Mod_2$) gate are equivalent up to constant depth, i.e., each can be
simulated by a constant-depth circuit using the other.  This is
completely different from the classical case, where parity cannot be
computed even with $\AC^0$ circuits, where fan-out is unrestricted
\cite{Ajtai:AC0,FSS:AC0}.  Later, Green et al.\ showed that quantum
$\Mod_q$-gates are constant-depth equivalent for all $q>1$, and
are thus all equivalent to fan-out.  Thus, for any $q>1$,
\[ \QNC^0_f = \QACC^0(q) = \QACC^0. \]
(The $f$ subscript means, ``with fan-out.'')  The classical analogs of
these classes are provably different.  In particular, classical
$\Mod_p$ and $\Mod_q$ gates are not constant-depth equivalent if $p$
and $q$ are distinct primes, and neither can be simulated by $\AC^0$
circuits \cite{Razborov:ACC,Smolensky:ACC}.

Using $\QNC^0$ circuits with unbounded fan-out gates, H{\o}yer \&
\v{S}palek managed to parallelize a sequence of commuting gates
applied to the same qubits, and thus greatly reduced the depth of
circuits for various purposes \cite{HS:fanout}.  They showed that
threshold gates can be approximated in constant depth this way, and
they can be computed exactly if Toffoli gates are also allowed.  Thus
$\QTC^0_f = \QACC^0$ as well.  Threshold gates, and hence fanout
gates, are quite powerful; many important arithmetic operations can be
computed in constant depth with threshold gates \cite{SBKH:threshold}.
This implies that the quantum Fourier transform---the quantum part of
Shor's factoring algorithm---can be approxmated in constant depth
using fanout gates.

All these results rely for their practicality on unbounded-width
quantum gates being available, especially fan-out or some (any) Mod
gate.  Unfortunately, making such a gate in the lab remains a daunting
prospect; it is hard enough just to fabricate a reliable CNOT gate.
Much more likely in the short term is that only one- and two-qubit
gates will be available, which brings us back to the now more
interesting question of $\QNC^0$.  How powerful is this class?  Can
$\QNC^0$ circuits be simulated classically, say, by computing their
acceptance probabilities either exactly or approximately?  Is there
\emph{anything} that $\QNC^0$ circuits can compute that cannot be
computed in classical polynomial time?  The present paper addresses
these questions.

A handful of hardness results about simulating constant-depth quantum
circuits with constant-width gates were given recently by Terhal \&
DiVincenzo \cite{TD:constant-depth}.  They showed that if one can
classically efficiently simulate, via sampling, the acceptance
probability of quantum circuits of depth at least three using one- and
two-qubit gates, then $\BQP \subseteq \AM$.  They also showed that the
polynomial hierarchy collapses if one can efficiently compute the
acceptance probability exactly for such circuits.  (Actually, a much
strong result follows from their proof, namely, $\P = \PP$.)  Their
technique uses an idea of Gottesman \& Chuang for teleporting CNOT
gates \cite{GC:teleportation} to transform an arbitrary quantum
circuit with CNOT and single-qubit gates into a depth-three circuit
whose acceptance probability is proportional to, though exponentially
smaller than, the original circuit.  Their results, however, only hold
on the supposition that depth-three circuits with \emph{arbitrary}
single-qubit and CNOT gates are simulatable.  We build on their
techniques, making improvements and simplifications.  We weaken their
hypothesis by showing how to produce a depth-three circuit with
essentially the same gates as the original circuit.  In addition, we
can get by with only with simple qubit state teleportation
\cite{BBCJPW:teleportation}.  Our results immediately show that the
class $\NQNC^0$ (the constant-depth analog of $\NQP$, see below), is
actually the same as $\NQP$, which is known to be as hard as the
polynomial hierarchy \cite{FGHP:NQP}.  We give this result in
Section~\ref{sec:main-lower-bound}.  It underscores yet another
drastic difference between the quantum and classical case: while
$\AC^0$ is well contained in $\P$, \ $\QNC^0$ circuits (even just
depth-three) can have amazingly complex behavior.  Our result is also
tight; Terhal \& DiVincenzo showed that the acceptance probabilities
of depth-two circuits over one- and two-qubit gates are computable in
polynomial time.

In Section~\ref{sec:main-upper-bound}, we give contrasting upper
bounds for $\QNC^0$-related language classes.  We show that various
bounded-error versions of $\QNC^0$ (defined below) are contained in
$\P$.  Particularly, $\EQNC^0 \subseteq \P$, where $\EQNC^0$ is the
constant-depth analog of the class $\EQP$ (see below).  Our proof uses
elementary probability theory, together with the fact that single
output qubit measurement probabilities can be computed directly, and
the fact that output qubits are ``largely'' independent of each other.
In hindsight, it is not too surprising that $\EQNC^0\subseteq\P$.
$\EQNC^0$ sets a severe limitation on the behavior of the circuit: it
must accept with certainty or reject with certainty.  This containment
is more surprising (to us) for the bounded-error $\QNC^0$ classes.

We give open questions and suggestions for further research in
Section~\ref{sec:open}.


\section{Preliminaries}
\label{sec:prelims}

\subsection{Gates and circuits}
\label{sec:prelims-gates}

We assume prior knowledge of basic concepts in computational
complexity: polynomial time, $\P$, $\NP$, as well as the counting
class $\NumP$ \cite{Valiant:NumP}.  Information can be found, for
example, in Papadimitriou \cite{Papadimitriou:complexity}.  The class
$\CneqP$ ($\co{\CeqP}$) was defined by Wagner \cite{Wagner:C=P}.  One
way of defining $\CneqP$ is as follows: a language $L$ is in $\CneqP$
iff there are two $\#P$ functions $f$ and $g$ such that, for all $x$,
\ $x\in L \iff f(x)\neq g(x)$.  $\CneqP$ was shown to be hard for the
polynomial hierarchy by Toda \& Ogihara \cite{TO:PHinBPGapP}.

We will also assume some (but less) background in quantum computation
and the quantum circuit model.  See Nielsen and Chuang
\cite{NC:quantumbook} for a good reference of basic concepts and
notation.

We review some standard quantum (unitary) gates.  Among the
single-qubit gates, we have the Pauli gates $X$, $Y$, and $Z$, the
Hadamard gate $H$, and the $\pi/8$ gate $T$, which are defined thus,
for $b\in\two$:
\begin{eqnarray*}
X\ket{b} & = & \ket{\neg b}, \\
Y\ket{b} & = & i(-1)^b\ket{\neg b}, \\
Z\ket{b} & = & (-1)^b\ket{b}, \\
H\ket{b} & = & (\ket{0} + (-1)^b\ket{1})/\sqrt{2}, \\
T\ket{b} & = & e^{i\pi b/4}\ket{b}.
\end{eqnarray*}
For $n \geq 1$, the $(n+1)$-qubit \emph{generalized Toffoli gate}
$T_n$ satisfies
\[ T_n\ket{x_1,\ldots,x_n,b} = \ket{x_1,\ldots,x_n,b\oplus
\bigwedge_{i=1}^n x_i}. \]
Here $b$ is the \emph{target qubit} and $x_1,\ldots,x_n$ are the
\emph{control qubits}.  $T_n$ is a kind of multiply controlled
$X$-gate (or NOT-gate), and is the quantum analog of the Boolean
AND-gate with fanin $n$.  $T_2$ is known simply as the Toffoli gate.
$T_1$ is also known as the controlled NOT (CNOT) gate and is depicted
below.  Here, $a,b\in\two$.
\begin{center}
\begin{picture}(0,0)%
\includegraphics{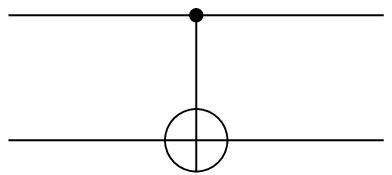}%
\end{picture}%
\setlength{\unitlength}{3947sp}%
\begingroup\makeatletter\ifx\SetFigFont\undefined%
\gdef\SetFigFont#1#2#3#4#5{%
  \reset@font\fontsize{#1}{#2pt}%
  \fontfamily{#3}\fontseries{#4}\fontshape{#5}%
  \selectfont}%
\fi\endgroup%
\begin{picture}(2100,852)(2251,-1723)
\put(4351,-1036){\makebox(0,0)[lb]{\smash{\SetFigFont{12}{14.4}{\rmdefault}{\mddefault}{\updefault}$a$}}}
\put(4351,-1636){\makebox(0,0)[lb]{\smash{\SetFigFont{12}{14.4}{\rmdefault}{\mddefault}{\updefault}$a\oplus b$}}}
\put(2251,-1036){\makebox(0,0)[rb]{\smash{\SetFigFont{12}{14.4}{\rmdefault}{\mddefault}{\updefault}$a$}}}
\put(2251,-1636){\makebox(0,0)[rb]{\smash{\SetFigFont{12}{14.4}{\rmdefault}{\mddefault}{\updefault}$b$}}}
\end{picture}
\end{center}

A gate closely related to $T_n$ is the controlled $Z$-gate defined by
\[ Z_n\ket{x_1,\ldots,x_n} = (-1)^{\bigwedge_{i=1}^n
x_i}\ket{x_1,\ldots,x_n}. \]
Since $HXH = Z$, the gate $Z_{n+1}$-gate can be implemented by placing
$H$-gates on either side of a $T_n$ gate on its target qubit.

The $(n+1)$-qubit fan-out gate $F_n$ is defined as follows:
\[ F_n\ket{x_1,\ldots,x_n,b} = \ket{x_1\oplus b,\ldots,x_n\oplus
b,b}. \]

For $q>1$, the $(n+1)$-qubit $\Mod_q$-gate acts on a basis state
$\ket{x_1,\ldots,x_n,b}$ by flipping the target qubit $b$ iff
$x_1+\cdots +x_n \not\equiv 0 \pmod{q}$.  The control qubits
$x_1,\ldots,x_n$ are left alone.  The $\Mod_2$ gate is also known as
the parity gate.  The \emph{width} of a gate is the number of qubits
on which it acts.

Our notion of quantum circuits is fairly standard (again see, for
example, \cite{NC:quantumbook}): a series of quantum gates, drawn from
some specified set of unitary operators, acting on some specified
number of qubits, labeled $1,\ldots,m$.  The first few qubits are
considered \emph{input} qubits, which are assumed to be in some basis
state initially (i.e., classical input); the rest are ancill\ae, each
assumed to be in the $\ket{0}$ state initially.  Thus the initial
state of the qubits is $\ket{x,00\cdots 0}$, for some binary string
$x$.  Some arbitrary set of qubits are specified as \emph{output}
qubits, and these qubits are measured in the computational basis at
the final state.  We assume that the sets of input and output qubits
are part of the description of the circuit.  The circuit
\emph{accepts} its input if all the output qubits are observed to be 0
in the final state.  Otherwise the circuit rejects.  We let
$\Pr[C(x)]$ denote the probability that $C$ accepts input $x$.

If $C$ is any quantum circuit, it will be convenient for us to define
$|C|$, the \emph{size} of $C$, to be the number of output qubits plus
the number of ``contact points'' between qubits and gates, so for
example, a single-qubit gate counts one towards the size, while a
two-qubit gate counts two, etc.  $C$ may be laid out by partitioning
its gates into \emph{layers} $1,\ldots,d$, such that (i) gates in the
same layer all act on pairwise disjoint sets of qubits, and (ii) all
gates in layer $i$ are applied before any gates in layer $i+1$, for
$1\leq i < d$.  The \emph{depth} of $C$ is then the smallest possible
value of $d$.  The \emph{width} of $C$ is the number of qubits in $C$.

The standard quantum complexity classes can be defined in terms of
quantum circuit families.  A quantum circuit family is a sequence
$\{C_n\}_{n\geq 0}$ of quantum circuits, where each $C_n$ has $n$
inputs.  We say that $\{C_n\}$ is \emph{uniform} if there is a
(classical) polynomial-time algorithm that outputs a description of
$C_n$ on input $0^n$.

\begin{definition}[\cite{BV:quantum,BBBV:quantum,ADH:quantum}]
Let $L$ be a language.
\begin{itemize}
\item
$L\in\EQP$ iff there is a uniform quantum circuit family $\{C_n\}$
such that, for all $x$ of length $n$,
\begin{eqnarray*}
x\in L & \implies & \Pr[C_n(x)] = 1, \\
x\notin L & \implies & \Pr[C_n(x)] = 0.
\end{eqnarray*}
\item
$L\in\BQP$ iff there is a uniform quantum circuit family $\{C_n\}$
such that, for all $x$ of length $n$,
\begin{eqnarray*}
x\in L & \implies & \Pr[C_n(x)] \geq 2/3, \\
x\notin L & \implies & \Pr[C_n(x)] < 1/3.
\end{eqnarray*}
\item
$L\in\NQP$ iff there is a uniform quantum circuit family $\{C_n\}$
such that, for all $x$ of length $n$,
\begin{eqnarray*}
x\in L & \implies & \Pr[C_n(x)] > 0, \\
x\notin L & \implies & \Pr[C_n(x)] = 0.
\end{eqnarray*}
\end{itemize}
\end{definition}

It is known that $\P \subseteq \EQP \subseteq \BQP$.  It was shown in
\cite{FGHP:NQP,YY:NQP} that $\NQP = \CneqP$, and is thus hard for the
polynomial hierarchy.

\subsection{Complexity classes using $\QNC$ circuits}

The circuit class $\QNC$ was first suggested by Moore and Nilsson
\cite{MN:quantum} as the quantum analog of the class $\NC$ of bounded
fan-in Boolean circuits with polylogarithmic depth and polynomial
size.  We define the class $\QNC^k$ in the same fashion as definitions
in Green, Homer, Moore, \& Pollett \cite{GHMP:QAC} with some minor
modifications.

\begin{definition}[\cite{MN:quantum}]
$\QNC^k$ is the class of quantum circuit families $\{C_n\}_{n\geq 0}$
for which there exists a polynomial $p$ such that each $C_n$ contains
$n$ input qubits and at most $p(n)$ many ancill\ae.  Each $C_n$ has
depth $O(\log^k n)$ and uses only single-qubit gates and CNOT gates.
The single-qubit gates must be from a fixed finite set.
\end{definition}

Next we define the language classes $\NQNC^k$ and $\EQNC^k$.  These
are $\QNC^k$ analogs of the classes $\NQP$ and $\EQP$, respectively.

\begin{definition}[\cite{GHMP:QAC}]
Let $k\geq 0$ be an integer.
\begin{itemize}
\item
$\NQNC^k$ is the class of languages $L$ such that there is a uniform
$\{C_n\}\in\QNC^k$ such that, for all $x$,
\[ x\in L \iff \Pr[C_{|x|}(x)]>0. \]
\item
$\EQNC^k$ is the class of languages $L$ such that there is a uniform
$\{C_n\}\in\QNC^k$ such that, for all $x$, \
$\Pr[C_{|x|}(x)]\in\{0,1\}$ and
\[ x\in L \iff \Pr[C_{|x|}(x)] = 1. \]
\end{itemize}
\end{definition}

\paragraph{Remark.}
Green, Homer, Moore, \& Pollett implicitly consider the output qubits
of $C_n$ to be all the qubits in $C_n$ \cite{GHMP:QAC}.  In our model
we allow any subset of qubits to be the output qubits of $C_n$, and we
do not restrict our circuits to be clean, i.e., the non-output qubits
could end up in an arbitrary state, possibly entangled with the output
qubits.  The reason we define our circuits this way is based on the
observation that, in their model, if a language $L$ is in $\EQNC^k$
(or $\BQNC^k_{\epsilon,\delta}$ for large enough $\delta$), then $L$
can contain no more than one string of each length.

\bigskip

Bounded-error $\QAC^k$ classes were mentioned in \cite{GHMP:QAC}, and
one can certainly ask about similar classes for $\QNC^k$ circuits.  It
is not obvious that there is one robust definition of
$\BQNC^0$---perhaps because it is not clear how to reduce error
significantly by amplification in constant depth.\footnote{One can
always reduce error classically by just running the circuit several
times on the same input.  In this case, the best definition of
$\BQNC^0$ may be that the gap between the allowed accept and reject
probabilities should be at least $1/{\rm poly}$.}  In the next
definition, we will try to be as general as possible while still
maintaining our assumption that $\vec{0}$ is the only accepting
output.

\begin{definition}
Let $\epsilon$ and $\delta$ be functions mapping (descriptions of)
quantum circuits into real numbers such that, for all quantum circuits
$C$, \ $0 < \epsilon(C) \leq \delta(C) \leq 1$.  We write
$\epsilon_C$ and $\delta_C$ to denote $\epsilon(C)$ and $\delta(C)$,
respectively.  $\BQNC^k_{\epsilon,\delta}$ is the class of languages
$L$ such that there is a uniform $\{C_n\}\in\QNC^k$ such that for any
string $x$ of length $n$,
\begin{eqnarray*}
x\in L & \implies & \Pr[C_n(x)]  \geq \delta_{C_n}, \\
x\notin L & \implies & \Pr[C_n(x)]  < \epsilon_{C_n}. 
\end{eqnarray*} 
\end{definition}

An interesting special case is when $\epsilon_C = \delta_C = 1$, that
is, the input is accepted iff the circuit accepts with probability 1,
and there is no promise on the acceptance probability.  One might
expect that, by the symmetry of the definitions, this class
$\BQNC^0_{1,1}$ is the same as $\NQNC^0$, but it is almost certainly
not, as we will see.

\subsection{Other classes of constant-depth quantum circuits}

\begin{definition}
Let $k \geq 0$ and $q > 1$ be integers.
\begin{itemize}
\item
$\QAC^k$ is the same as $\QNC^k$ except that generalized Toffoli gates
are allowed in the circuits.
\item
$\QACC(q)$ is the same as $\QNC^0$ except that $\Mod_q$ gates are
allowed in the circuits.
\item
$\QACC = \bigcup_{q>1} \QACC(q)$.
\end{itemize}
\end{definition}

\section{Main results}

\subsection{Simulating $\QNC^0$ circuits exactly is hard}
\label{sec:main-lower-bound}

\begin{theorem}\label{thm:lower-bound}
$\NQNC^0 = \NQP = \CneqP$.
\end{theorem}

As a corollary, we essentially solve an open problem of Green et al.\
\cite{GHMP:QAC}.  They conjectured that $\NQACC \subseteq \TC^0$, the
class of constant-depth Boolean circuits with threshold gates.

\begin{corollary}
For any $k\geq 0$,
\[ \NQNC^0 = \NQNC^k = \NQAC^k = \NQACC = \CneqP. \]
Thus, $\NQACC \not\subseteq \TC^0$ unless $\CneqP = \TC^0$.
\end{corollary}

Let $B$ be the two-qubit Bell gate, defined as
\begin{center}
\begin{picture}(0,0)%
\includegraphics{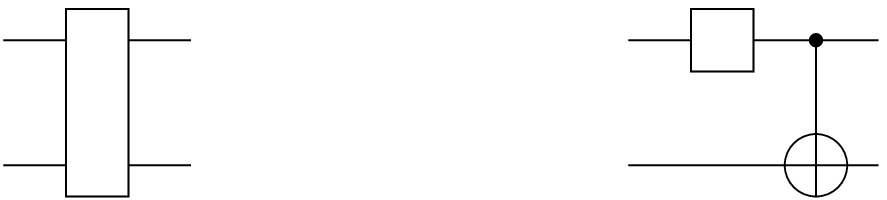}%
\end{picture}%
\setlength{\unitlength}{3947sp}%
\begingroup\makeatletter\ifx\SetFigFont\undefined%
\gdef\SetFigFont#1#2#3#4#5{%
  \reset@font\fontsize{#1}{#2pt}%
  \fontfamily{#3}\fontseries{#4}\fontshape{#5}%
  \selectfont}%
\fi\endgroup%
\begin{picture}(4224,924)(1789,-1723)
\put(3751,-1336){\makebox(0,0)[b]{\smash{\SetFigFont{12}{14.4}{\rmdefault}{\mddefault}{\updefault}$:=$}}}
\put(5251,-1036){\makebox(0,0)[b]{\smash{\SetFigFont{12}{14.4}{\rmdefault}{\mddefault}{\updefault}$H$}}}
\put(2251,-1336){\makebox(0,0)[b]{\smash{\SetFigFont{12}{14.4}{\rmdefault}{\mddefault}{\updefault}$B$}}}
\end{picture}
\end{center}
Also let
\begin{center}
\begin{picture}(0,0)%
\includegraphics{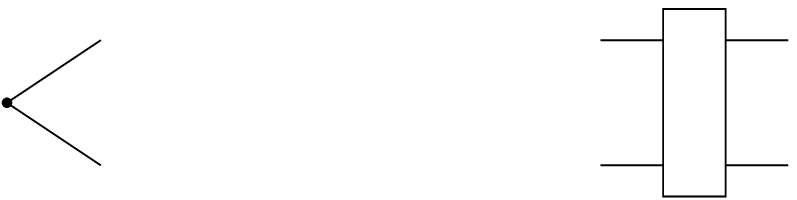}%
\end{picture}%
\setlength{\unitlength}{3947sp}%
\begingroup\makeatletter\ifx\SetFigFont\undefined%
\gdef\SetFigFont#1#2#3#4#5{%
  \reset@font\fontsize{#1}{#2pt}%
  \fontfamily{#3}\fontseries{#4}\fontshape{#5}%
  \selectfont}%
\fi\endgroup%
\begin{picture}(3791,924)(2222,-1723)
\put(5026,-1036){\makebox(0,0)[rb]{\smash{\SetFigFont{12}{14.4}{\rmdefault}{\mddefault}{\updefault}$\ket{0}$}}}
\put(5026,-1636){\makebox(0,0)[rb]{\smash{\SetFigFont{12}{14.4}{\rmdefault}{\mddefault}{\updefault}$\ket{0}$}}}
\put(3901,-1336){\makebox(0,0)[b]{\smash{\SetFigFont{12}{14.4}{\rmdefault}{\mddefault}{\updefault}$:=$}}}
\put(5551,-1336){\makebox(0,0)[b]{\smash{\SetFigFont{12}{14.4}{\rmdefault}{\mddefault}{\updefault}$B$}}}
\end{picture}
\end{center}
which produces the EPR state $(\ket{00} + \ket{11})/\sqrt{2}$.
We prove the following lemma, from which the theorem follows quickly.

\begin{lemma}
For any quantum circuit $\cC$ using gates drawn from any family ${\cal
F}$, there is a depth-three quantum circuit $\cC'$ of size linear in
$|\cC|$ using gates drawn from ${\cal F}\cup\{B,\adj{B}\}$ such that
for any input $x$ of the appropriate length,
\[ \Pr[\cC'(x)] = 2^{-m} \Pr[\cC(x)], \]
for some $m \leq 2|\cC|$ depending only on $\cC$.  The middle layer
of $\cC'$ contains each gate in $\cC$ exactly once and no others.  The
third layer contains only $\adj{B}$-gates, and the first layer
contains only $B$-gates, which are used only to create EPR states.
\end{lemma}

\begin{proof}
Our construction is a simplified version of the main construction in
Terhal \& DiVincenzo \cite{TD:constant-depth}, but ours is stronger
in one crucial respect discussed below: it does not significantly
increase the family of gates used.  To construct $\cC'$, we start with
$\cC$ and simply insert, for each qubit $q$ of $\cC$, a simplified
teleportation module (shown in Figure~\ref{fig:teleport})
\begin{figure}
\begin{center}
\begin{picture}(0,0)%
\includegraphics{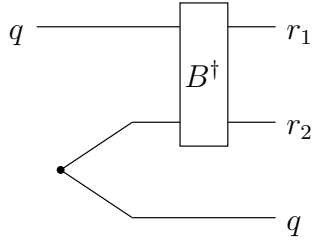}%
\end{picture}%
\setlength{\unitlength}{3947sp}%
\begingroup\makeatletter\ifx\SetFigFont\undefined%
\gdef\SetFigFont#1#2#3#4#5{%
  \reset@font\fontsize{#1}{#2pt}%
  \fontfamily{#3}\fontseries{#4}\fontshape{#5}%
  \selectfont}%
\fi\endgroup%
\begin{picture}(1650,1495)(4426,-2294)
\put(4426,-1036){\makebox(0,0)[rb]{\smash{\SetFigFont{12}{14.4}{\rmdefault}{\mddefault}{\updefault}$q$}}}
\put(6076,-2236){\makebox(0,0)[lb]{\smash{\SetFigFont{12}{14.4}{\rmdefault}{\mddefault}{\updefault}$q$}}}
\put(6076,-1036){\makebox(0,0)[lb]{\smash{\SetFigFont{12}{14.4}{\rmdefault}{\mddefault}{\updefault}$r_1$}}}
\put(6076,-1636){\makebox(0,0)[lb]{\smash{\SetFigFont{12}{14.4}{\rmdefault}{\mddefault}{\updefault}$r_2$}}}
\put(5551,-1336){\makebox(0,0)[b]{\smash{\SetFigFont{12}{14.4}{\rmdefault}{\mddefault}{\updefault}$\adj{B}$}}}
\end{picture}
\caption{The nonadaptive teleportation module
\cite{TD:constant-depth}.  The state in qubit $q$ is teleported
correctly iff the qubits $r_1$ and $r_2$ are both observed to be 0.}
\label{fig:teleport}
\end{center}
\end{figure}
between any two consecutive quantum gates of $\cC$ acting on $q$.  No
further gates involve the qubits $r_1$ and $r_2$ to the right of the
$\adj{B}$-gate.  This module, which lacks the usual corrective Pauli
gates, is a nonadaptive version of the standard single-qubit
teleportation circuit \cite{BBCJPW:teleportation}.  It faithfully
teleports the state if and only if the observed output of the
$\adj{B}$-gate on the right is $00$.
After inserting each teleportation circuit, the gates acting before
and after it are now acting on different qubits.  Further, it is
important to note that any entanglement the qubit state has with other
qubits is easily seen to be preserved in the teleported qubit.  The
input qubits of $\cC'$ are those of $\cC$.  The output qubits of
$\cC'$ are of two kinds: output qubits corresponding to outputs of
$\cC$ are the \emph{original outputs}; the other outputs are the
qubits (in pairs) coming from the added $\adj{B}$-gates.  We'll call
the measurement of each such pair a \emph{Bell measurement},
even though it is really in the computational basis.

In addition to the gates in $\cC$, \ $\cC'$ uses only $B$-gates to
make the initial EPR pairs and $\adj{B}$-gates for the Bell
measurements.  A sample transformation is shown in
Figure~\ref{fig:sample-trans}.
\begin{figure}
\begin{center}
\begin{picture}(0,0)%
\includegraphics{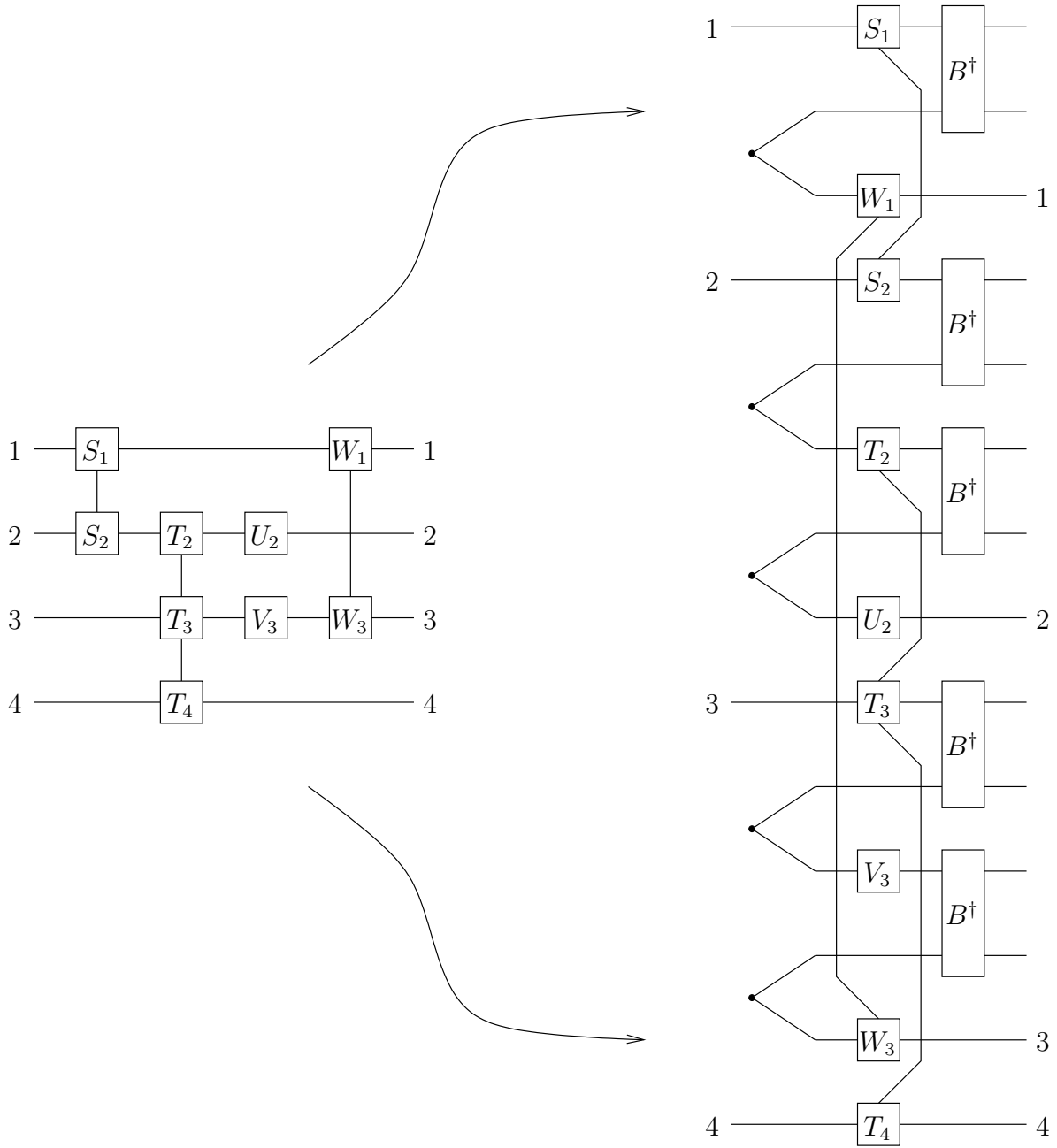}%
\end{picture}%
\setlength{\unitlength}{3947sp}%
\begingroup\makeatletter\ifx\SetFigFont\undefined%
\gdef\SetFigFont#1#2#3#4#5{%
  \reset@font\fontsize{#1}{#2pt}%
  \fontfamily{#3}\fontseries{#4}\fontshape{#5}%
  \selectfont}%
\fi\endgroup%
\begin{picture}(7200,8124)(976,-8623)
\put(8176,-1936){\makebox(0,0)[lb]{\smash{\SetFigFont{12}{14.4}{\rmdefault}{\mddefault}{\updefault}1}}}
\put(8176,-4936){\makebox(0,0)[lb]{\smash{\SetFigFont{12}{14.4}{\rmdefault}{\mddefault}{\updefault}2}}}
\put(8176,-7936){\makebox(0,0)[lb]{\smash{\SetFigFont{12}{14.4}{\rmdefault}{\mddefault}{\updefault}3}}}
\put(8176,-8536){\makebox(0,0)[lb]{\smash{\SetFigFont{12}{14.4}{\rmdefault}{\mddefault}{\updefault}4}}}
\put(5926,-736){\makebox(0,0)[rb]{\smash{\SetFigFont{12}{14.4}{\rmdefault}{\mddefault}{\updefault}1}}}
\put(5926,-2536){\makebox(0,0)[rb]{\smash{\SetFigFont{12}{14.4}{\rmdefault}{\mddefault}{\updefault}2}}}
\put(5926,-5536){\makebox(0,0)[rb]{\smash{\SetFigFont{12}{14.4}{\rmdefault}{\mddefault}{\updefault}3}}}
\put(5926,-8536){\makebox(0,0)[rb]{\smash{\SetFigFont{12}{14.4}{\rmdefault}{\mddefault}{\updefault}4}}}
\put(976,-3736){\makebox(0,0)[rb]{\smash{\SetFigFont{12}{14.4}{\rmdefault}{\mddefault}{\updefault}1}}}
\put(976,-4336){\makebox(0,0)[rb]{\smash{\SetFigFont{12}{14.4}{\rmdefault}{\mddefault}{\updefault}2}}}
\put(976,-4936){\makebox(0,0)[rb]{\smash{\SetFigFont{12}{14.4}{\rmdefault}{\mddefault}{\updefault}3}}}
\put(976,-5536){\makebox(0,0)[rb]{\smash{\SetFigFont{12}{14.4}{\rmdefault}{\mddefault}{\updefault}4}}}
\put(3826,-3736){\makebox(0,0)[lb]{\smash{\SetFigFont{12}{14.4}{\rmdefault}{\mddefault}{\updefault}1}}}
\put(3826,-4336){\makebox(0,0)[lb]{\smash{\SetFigFont{12}{14.4}{\rmdefault}{\mddefault}{\updefault}2}}}
\put(3826,-4936){\makebox(0,0)[lb]{\smash{\SetFigFont{12}{14.4}{\rmdefault}{\mddefault}{\updefault}3}}}
\put(3826,-5536){\makebox(0,0)[lb]{\smash{\SetFigFont{12}{14.4}{\rmdefault}{\mddefault}{\updefault}4}}}
\put(7051,-736){\makebox(0,0)[b]{\smash{\SetFigFont{12}{14.4}{\rmdefault}{\mddefault}{\updefault}$S_1$}}}
\put(7651,-1036){\makebox(0,0)[b]{\smash{\SetFigFont{12}{14.4}{\rmdefault}{\mddefault}{\updefault}$\adj{B}$}}}
\put(7051,-2536){\makebox(0,0)[b]{\smash{\SetFigFont{12}{14.4}{\rmdefault}{\mddefault}{\updefault}$S_2$}}}
\put(7051,-1936){\makebox(0,0)[b]{\smash{\SetFigFont{12}{14.4}{\rmdefault}{\mddefault}{\updefault}$W_1$}}}
\put(7651,-2836){\makebox(0,0)[b]{\smash{\SetFigFont{12}{14.4}{\rmdefault}{\mddefault}{\updefault}$\adj{B}$}}}
\put(7051,-3736){\makebox(0,0)[b]{\smash{\SetFigFont{12}{14.4}{\rmdefault}{\mddefault}{\updefault}$T_2$}}}
\put(7651,-4036){\makebox(0,0)[b]{\smash{\SetFigFont{12}{14.4}{\rmdefault}{\mddefault}{\updefault}$\adj{B}$}}}
\put(7051,-4936){\makebox(0,0)[b]{\smash{\SetFigFont{12}{14.4}{\rmdefault}{\mddefault}{\updefault}$U_2$}}}
\put(7051,-5536){\makebox(0,0)[b]{\smash{\SetFigFont{12}{14.4}{\rmdefault}{\mddefault}{\updefault}$T_3$}}}
\put(7051,-6736){\makebox(0,0)[b]{\smash{\SetFigFont{12}{14.4}{\rmdefault}{\mddefault}{\updefault}$V_3$}}}
\put(7651,-5836){\makebox(0,0)[b]{\smash{\SetFigFont{12}{14.4}{\rmdefault}{\mddefault}{\updefault}$\adj{B}$}}}
\put(7651,-7036){\makebox(0,0)[b]{\smash{\SetFigFont{12}{14.4}{\rmdefault}{\mddefault}{\updefault}$\adj{B}$}}}
\put(7051,-7936){\makebox(0,0)[b]{\smash{\SetFigFont{12}{14.4}{\rmdefault}{\mddefault}{\updefault}$W_3$}}}
\put(7051,-8536){\makebox(0,0)[b]{\smash{\SetFigFont{12}{14.4}{\rmdefault}{\mddefault}{\updefault}$T_4$}}}
\put(2101,-4336){\makebox(0,0)[b]{\smash{\SetFigFont{12}{14.4}{\rmdefault}{\mddefault}{\updefault}$T_2$}}}
\put(2701,-4336){\makebox(0,0)[b]{\smash{\SetFigFont{12}{14.4}{\rmdefault}{\mddefault}{\updefault}$U_2$}}}
\put(2101,-4936){\makebox(0,0)[b]{\smash{\SetFigFont{12}{14.4}{\rmdefault}{\mddefault}{\updefault}$T_3$}}}
\put(2101,-5536){\makebox(0,0)[b]{\smash{\SetFigFont{12}{14.4}{\rmdefault}{\mddefault}{\updefault}$T_4$}}}
\put(3301,-3736){\makebox(0,0)[b]{\smash{\SetFigFont{12}{14.4}{\rmdefault}{\mddefault}{\updefault}$W_1$}}}
\put(3301,-4936){\makebox(0,0)[b]{\smash{\SetFigFont{12}{14.4}{\rmdefault}{\mddefault}{\updefault}$W_3$}}}
\put(2701,-4936){\makebox(0,0)[b]{\smash{\SetFigFont{12}{14.4}{\rmdefault}{\mddefault}{\updefault}$V_3$}}}
\put(1501,-3736){\makebox(0,0)[b]{\smash{\SetFigFont{12}{14.4}{\rmdefault}{\mddefault}{\updefault}$S_1$}}}
\put(1501,-4336){\makebox(0,0)[b]{\smash{\SetFigFont{12}{14.4}{\rmdefault}{\mddefault}{\updefault}$S_2$}}}
\end{picture}
\caption{A sample transformation from $\cC$ to $\cC'$.  The circuit
$\cC$ on the left has five gates: $S$, $T$, $U$, $V$, and $W$, with
subscripts added to mark which qubits each gate is applied to.  The
qubits in $\cC'$ are numbered corresponding to those in $\cC$.}
\label{fig:sample-trans}
\end{center}
\end{figure}
$\cC'$ has depth three since it uses the first layer to make the
initial EPR states and the third layer to rotate the Bell basis back
to the computational basis.  All the gates of $\cC$ appear on the
second layer.  From the above constuction and the properties of the
teleportation module, it is not hard to see that for all $x$ of the
appropriate length,
\begin{eqnarray*}
\Pr[\cC(x)] & = & \Pr[\mbox{all original outputs of $\cC'$ are 0} \mid
\mbox{all qubit states are teleported correctly}] \\
& = & \Pr[\mbox{all original outputs of are 0} \mid \mbox{all
Bell measurement results are 00}] \\
& = & \frac{\Pr[\cC'(x)]}{\Pr[\mbox{all Bell measurement results are
00}]},
\end{eqnarray*}
since the Bell measurements are among the output measurements of
$\cC'$.  Let $k$ be the number of $\adj{B}$-gates on layer 3.
Clearly, $k \leq |\cC|$, and it is well-known that each Bell
measurement will give 00 with probability $1/4$, independent of all
other measurements.  So the lemma follows by setting $m = 2k$.
\end{proof}

\begin{proofof}{Theorem~\ref{thm:lower-bound}}
As mentioned before, $\NQP$ \cite{ADH:quantum} is defined as the class
of languages recognized by quantum Turing machines (equivalently,
uniform quantum circuit families over a finite set of gates) where the
acceptance criterion is that the accepting state appear with nonzero
probability.  It is known \cite{FGHP:NQP,YY:NQP} that $\NQP = \CneqP$,
which contains $\NP$ and is hard for the polynomial hierarchy.  Since
$\QNC^0$ circuit families must also draw their gates from some finite
set, we clearly have $\NQNC^0 \subseteq \NQP$.  The reverse
containment follows from our construction: an arbitrary circuit $\cC$
is transformed into a depth-three circuit $\cC'$ \emph{with the same
gates} as $\cC$ plus $B$ and $\adj{B}$.  Moreover, $\cC'$ accepts with
nonzero probability iff $\cC$ does.  Thus an $\NQP$ language $L$
recognized by a uniform quantum circuit family over a finite set of
quantum gates is also recognized by a uniform depth-three circuit
family over a finite set of quantum gates, and so $L\in\NQNC^0$.
\end{proofof}

Using the gate teleportation apparatus of Gottesmann and Chuang
\cite{GC:teleportation}, Terhal \& DiVincenzo also construct a
depth-three\footnote{They count the depth as four, but they include
the final measurement as an additional layer whereas we do not.}
quantum circuit $\cC'$ out of an arbitrary circuit $\cC$ (over CNOT
and single-qubit gates) with a similar relationship of acceptance
probabilities.  However, they only teleport the CNOT gate, and their
$\cC'$ may contain single-qubit gates formed by compositions of
arbitrary numbers of single-qubit gates from $\cC$.  (Such gates may
not even be approximable in constant depth by circuits over a fixed
finite family of gates.)  When their construction is applied to each
circuit in a uniform family, the resulting circuits are thus not
generally over a finite gate set, even if the original circuits were.

Our construction solves this problem by teleporting every qubit state
in between all gates involving it.  Besides $B$ and $\adj{B}$, we only
use the gates of the original circuit.  We also are able to bypass the
CNOT gate teleportation technique of \cite{GC:teleportation}, using
instead basic single-qubit teleportation \cite{BBCJPW:teleportation},
which works with arbitrary gates.

\subsection{Simulating $\QNC^0$ circuits approximately is easy}
\label{sec:main-upper-bound}

In this section we prove that $\BQNC^0_{\epsilon,\delta} \subseteq \P$
for certain $\epsilon,\delta$.  For convenience we will assume that
all gates used in quantum circuits are either one- or two-qubit gates
that have ``reasonable'' matrix elements---algebraic numbers, for
instance.  Our results can apply more broadly, but they will then
require greater care to prove.

For a quantum circuit $\cC$, we define a dependency graph over the set
of its output qubits.

\begin{definition}
Let $\cC$ be a quantum circuit and let $p$ and $q$ be qubits of
$\cC$.  We say that \emph{$q$ depends on $p$} if there is a forward
path in $\cC$ starting at $p$ before the first layer, possibly passing
through gates, and ending at $q$ after the last layer.  More formally,
we can define dependence by induction on the depth of $\cC$.  For depth
zero, $q$ depends on $p$ iff $q = p$.  For depth $d>0$, let $\cC'$ be
the same as $\cC$ but missing the first layer.  Then $q$ depends on
$p$ (in $\cC$) iff there is a qubit $r$ such that $q$ depends on $r$
(in $\cC'$) and either $p = r$ or there is a gate on the first layer
of $\cC$ that involves both $p$ and $r$.
\end{definition}

\begin{definition}
For $\cC$ a quantum circuit and $q$ a qubit of $\cC$, define
\[ D_q = \{ p \mid \mbox{$q$ depends on $p$}\}. \]
If $S$ is a set of qubits of $\cC$, define $D_S = \bigcup_{q\in S}
D_q$.  Let the \emph{dependency graph of $\cC$} be the undirected
graph with the output qubits of $\cC$ as vertices, and with an edge
between two qubits $q_1$ and $q_2$ iff $D_{q_1} \cap D_{q_2} \neq
\emptyset$.
\end{definition}

If $\cC$ has depth $d$, then it is easy to see that the degree of its
dependency graph is less than $2^{2d}$.  The following lemma is
straightforward.

\begin{lemma}\label{lem:independence}
Let $\cC$ be a quantum circuit and let $S$ and $T$ be sets of output
qubits of $\cC$.  Fix an input $x$ and bit vectors $u$ and $v$ with
lengths equal to the sizes of $S$ and $T$, respectively.  Let
$E_{S=u}$ (respectively $E_{T=v}$) be the event that the qubits in $S$
(respectively $T$) are observed to be in the state $u$ (respectively
$v$) in the final state of $\cC$ on input $x$.  If $D_S \cap D_T =
\emptyset$, then $E_{S=u}$ and $E_{T=v}$ are independent.
\end{lemma}

For an algebraic number $a$, we let $\|a\|$ be the size of some
reasonable representation of $a$.

The results in this section follow from the next theorem.

\begin{theorem}\label{thm:upper-bound}
There is a deterministic decision algorithm $A$ which takes as input
\begin{enumerate}
\item
a quantum circuit $\cC$ with depth $d$ and $n$ input qubits,
\item
a binary string $x$ of length $n$, and
\item
an algebraic number $t \in [0,1]$,
\end{enumerate}
and behaves as follows: Let $D$ be one plus the degree of the
dependency graph of $\cC$.  $A$ runs in time
$\Poly(|\cC|,2^{2^d},\|t\|)$, and
\begin{itemize}
\item
if $\Pr[\cC(x)] \geq 1-t$, then $A$ accepts, and
\item
if $\Pr[\cC(x)] < 1-Dt$, then $A$ rejects.
\end{itemize}
\end{theorem}

Note that since $D \leq 2^{2d}$, if $t < 2^{-2d}$, then $A$ will
reject when $\Pr[\cC(x)] < 1-2^{2d}t$.

\begin{proofof}{Theorem~\ref{thm:upper-bound}}
On input $(\cC,x,t)$ as above,
\begin{enumerate}
\item
$A$ computes the dependency graph $G = (V,E)$ of $\cC$ and its degree,
and sets $D$ to be the degree plus one.
\item
$A$ finds a $D$-coloring $\map{c}{V}{\{1,\ldots,D\}}$ of $G$ via a
standard greedy algorithm.
\item
For each output qubit $q\in V$, \ $A$ computes $P_q$---the probability
that 0 is measured on qubit $q$ in the final state (given input $x$).
\item
For each color $i\in\{1,\ldots,D\}$, let $B_i = \{ q\in V \mid c(q) =
i \}$.  $A$ computes
\[ P_{B_i} = \prod_{q\in B_i} P_q, \]
which by Lemma~\ref{lem:independence} is the probability that all
qubits colored $i$ are observed to be 0 in the final state.
\item
If $P_{B_i} \geq 1-t$ for all $i$, the $A$ accepts;
otherwise, $A$ rejects.
\end{enumerate}

We first show that $A$ is correct.  If $\Pr[\cC(x)] \geq 1-t$, then
for each $i\in\{1,\ldots,D\}$,
\[ 1-t \leq \Pr[\cC(x)] \leq P_{B_i}, \]
and so $A$ accepts.  On the other hand, if $\Pr[\cC(x)] < 1-Dt$, then
\[ Dt < 1 - \Pr[\cC(x)] \leq \sum_{i=1}^D \left(1-P_{B_i}\right), \]
so there must exist an $i$ such that $t < 1 - P_{B_i}$, and thus $A$
rejects.

To show that $A$ runs in the given time, first we show that the
measurement statistics of any output qubit can be calculated in time
polynomial in $2^{2^d}$.  Pick an output qubit $q$.  By looking at
$\cC$ we can find $D_q$ in time $\Poly(|\cC|)$.  Since $\cC$ has depth
$d$ and uses gates on at most two qubits each, $D_q$ had cardinality
at most $2^d$.  Then we simply calculate the measurement statistics of
output qubit $q$ from the input state restricted to $D_q$, i.e., with
the other qubits traced out.  This can be done by computing the state
layer by layer, starting with layer one, and at each layer tracing out
qubits when they no longer can reach $q$.  Because of the partial
traces, the state will in general be a mixed state so we maintain it
as a density operator.  We are multiplying matrices of size at most
$2^{2^d}\times 2^{2^d}$ at most $O(d)$ times.  All this will take time
polynomial in $2^{2^d}$, provided we can show that the individual
field operations on the matrix elements do not take too long.

Since there are finitely many gates to choose from, their (algebraic)
matrix elements generate a field extension $F$ of $\rats$ with finite
index $r$.  We can thus store values in $F$ as $r$-tuples of rational
numbers, with the field operations of $F$ taking polynomial time.
Furthermore, one can show that for $a,b\in F$, \ $\|ab\| = O(\|a\| +
\|b\|)$ and $\left\|\sum_{i=1}^n a_i\right\| = O(n \cdot \max_i
\|a_i\|)$ for any $a_1,\ldots,a_n\in F$.  A bit of calculation then
shows that the intermediate representations of numbers do not get too
large.

The dependency graph and its coloring can clearly be computed in time
$\Poly(|\cC|)$.  The only things left are the computation of the
$P_{B_i}$ and their comparison with $1-t$.  For reasons similar to
those above for matrix multiplication, this can be done in time
$\Poly(|\cC|,2^{2^d},\|t\|)$.
%
%
%
\end{proofof}

\begin{corollary}\label{cor:master}
Suppose $\epsilon$ and $\delta$ are polynomial-time computable, and for
any quantum circuit $C$ of depth $d$, \ $\delta_C = 1 -
2^{-2d}(1-\epsilon_C)$.  Then
\[ \BQNC^0_{\epsilon,\delta} \subseteq \P. \]
\end{corollary}

\begin{proof}
For each $C$ of depth $d$ in the circuit family and each input $x$,
apply the algorithm $A$ of Theorem~\ref{thm:upper-bound} with $t =
1-\delta_C = 2^{-2d}(1-\epsilon_C)$, noting that $D \leq 2^{2d}$.
\end{proof}

The following few corollaries are instances of
Corollary~\ref{cor:master}.

\begin{corollary}
For quantum circuit $C$, let
$\delta_C = 1 - 2^{-(2d+1)}$, where $d$ is the depth of $C$.  Then
\[ \BQNC^0_{(1/2),\delta} \subseteq \P. \]
\end{corollary}

\begin{proof}
Apply algorithm $A$ to each circuit, setting $t = 2^{-(2d+1)}$.
\end{proof}

\begin{corollary}\label{cor:BQNC0}
$\BQNC^0_{1,1} \subseteq \P$.
\end{corollary}

\begin{proof}
Apply algorithm $A$ to each circuit, setting $t = 0$.
\end{proof}

\begin{corollary}\label{cor:EQNC0}
$\EQNC^0 \subseteq \P$.
\end{corollary}

\begin{proof}
Clearly, $\EQNC^0 \subseteq \BQNC^0_{1,1}$.
\end{proof}

Corollaries~\ref{cor:BQNC0} and \ref{cor:EQNC0} can actually be proven
more directly.  We simply compute, for each output, its probability of
being 0.  We accept iff all probabilities are 1.


We observe here that by a simple proof using our techniques, one can
show that the generalized Toffoli gate cannot be simulated by a
$\QNC^0$ circuit, since the target of the Toffoli gate can only depend
on constantly may input qubits.

\section{Conclusions, open questions, and further research}
\label{sec:open}

Our upper bound results in Section~\ref{sec:main-upper-bound} can be
improved in certain ways.  For example, the containment in $\P$ is
easily seen to apply to $(\log\log n + O(1))$-depth circuits as well.
Can we increase the depth further?  Another improvement would be to
put $\BQNC^0_{\epsilon,\delta}$ into classes smaller than $\P$.
LOGSPACE seems managable.  How about $\NC^1$?

There are some general questions about whether we have the ``right''
definitions for these classes.  For example, the accepting outcome is
defined to be all outputs being 0.  One can imagine more general
accepting conditions, such as arbitrary classical polynomial-time
postprocessing.  If we allow this, then all our classes will obviously
contain $\P$.  If we allow arbitrary classical polynomial-time
\emph{pre}processing, then all our classes will be closed under ptime
$m$-reductions (Karp reductions).

Finally, there is the question of the probability gap in the
definitions of $\BQNC^0$.  Certainly we would like to narrow this gap
(ideally, to $1/{\rm poly}$) and still get containment in $\P$.

\section*{Acknowledgments}

We would like to thank David DiVincenzo and Mark Heiligman for helpful
conversations on this topic.  This work was supported in part by the
National Security Agency (NSA) and Advanced Research and Development
Agency (ARDA) under Army Research Office (ARO) contract numbers
DAAD~19-02-1-0058 (for S.~Homer, and F.~Green) and DAAD~19-02-1-0048
(for S.~Fenner and Y.~Zhang).

\bibliography{eqnc.bbl}

\end{document}